\documentclass[%
 reprint,
superscriptaddress,
 amsmath,amssymb,
 aps,
]{revtex4-2}

\usepackage{graphicx}
\usepackage{dcolumn}
\usepackage{bm}
\usepackage{hyperref}
\usepackage{upgreek}
\usepackage{amsmath}
\usepackage{siunitx}
\usepackage{tabularx}
\usepackage{makecell}

\usepackage{pgffor}

\newcommand{\bigconv}{\mathop{\scalebox{1.5}{\raisebox{-0.2ex}{$\ast$}}}}
\newcommand{\mat}[1]{\boldsymbol{#1}}

\begin{document}

\preprint{APS/123-QED}

\title{Macroscopic photon counting beating the Poisson noise limit}

\author{Timon Schapeler}
\email{timon.schapeler@upb.de}
\affiliation{Department of Physics, Paderborn University, Warburger Str. 100, 33098 Paderborn, Germany}
\affiliation{Institute for Photonic Quantum Systems (PhoQS), Paderborn University, Warburger Str. 100, 33098 Paderborn, Germany}

\author{Fabian Schlue}
\affiliation{Department of Physics, Paderborn University, Warburger Str. 100, 33098 Paderborn, Germany}
\affiliation{Institute for Photonic Quantum Systems (PhoQS), Paderborn University, Warburger Str. 100, 33098 Paderborn, Germany}

\author{Isabell Mischke}
\affiliation{Department of Physics, Paderborn University, Warburger Str. 100, 33098 Paderborn, Germany}
\affiliation{Institute for Photonic Quantum Systems (PhoQS), Paderborn University, Warburger Str. 100, 33098 Paderborn, Germany}

\author{Michael Stefszky}
\affiliation{Department of Physics, Paderborn University, Warburger Str. 100, 33098 Paderborn, Germany}
\affiliation{Institute for Photonic Quantum Systems (PhoQS), Paderborn University, Warburger Str. 100, 33098 Paderborn, Germany}

\author{Benjamin Brecht}
\affiliation{Department of Physics, Paderborn University, Warburger Str. 100, 33098 Paderborn, Germany}
\affiliation{Institute for Photonic Quantum Systems (PhoQS), Paderborn University, Warburger Str. 100, 33098 Paderborn, Germany}

\author{Christine Silberhorn}
\affiliation{Department of Physics, Paderborn University, Warburger Str. 100, 33098 Paderborn, Germany}
\affiliation{Institute for Photonic Quantum Systems (PhoQS), Paderborn University, Warburger Str. 100, 33098 Paderborn, Germany}

\author{Tim J. Bartley}
\affiliation{Department of Physics, Paderborn University, Warburger Str. 100, 33098 Paderborn, Germany}
\affiliation{Institute for Photonic Quantum Systems (PhoQS), Paderborn University, Warburger Str. 100, 33098 Paderborn, Germany}


\begin{abstract}
Photon counting is a cornerstone of quantum optics. Here, we demonstrate precisely counting from 0 to over 9000 photons, beating the Poisson noise limit by at least $4.1~\mathrm{dB}$ across this range. We achieve sub-single-photon precision up to 276 photons per pulse. To do so, we multiplex eight intrinsically photon-number-resolving superconducting nanowire single-photon detectors across 128 temporal modes. We use a model-informed characterization of each of the 1024 detection bins, for optimal precision. 
We perform quantum detector tomography to reconstruct the positive operator valued measures (POVMs) of the complete device, which consists of $1.38\cdot10^8$ matrix elements.
At the repetition rate of our experiment of $80~\mathrm{kHz}$, we can precisely count photons corresponding to an optical power of approximately $71~\mathrm{pW}$, bridging the gap from single-photon measurements to high-sensitivity optical power meters. A photon-number-resolving detector of this size, and the tools used to analyze it, will become increasingly important to characterize large quantum states, as well as tasks in precision metrology and optical power standards.
\end{abstract}

\maketitle

\section{Introduction}\label{sec:intro}
The quantization of the energy of a light wave is one of the most fundamental aspects of quantum mechanics. Nevertheless, building detectors which remain sensitive to the discrete nature of photons when the intensity of the light field is large is highly challenging. The most commonly used single-photon detectors (such as avalanche photodiodes, or superconducting nanowire single-photon detectors) typically distinguish only the presence or absence of photons, not their number~\cite{migdall2013single}. Some photon-number information can be recovered by multiplexing these devices: splitting the incident light across many detector elements or pixels~\cite{paul1996photon,achilles2003fiber,fitch2003photon,cheng2023a}. In the limit of equal splitting across infinite pixels, this recovers the photon number information precisely~\cite{sperling2012true}. In practice, finite pixel numbers and nonuniform splitting between them mean that there is non-zero probability of multiple photons incident on the same pixel, causing significant loss and statistical noise~\cite{sperling2012true}. Other types of detector, such as transition edge sensors (TESs), may resolve energy directly (corresponding to photon-number resolution if the incident light is monochromatic) but these devices become saturated beyond $\sim 20$ photons~\cite{lita2008counting,harder2016single-mode,eaton2023resolution}. TES can also be operated in the saturated regime, where photon-number information up to millions of photons is contained in their recovery time~\cite{gerrits2012extending}. Alternatively, detectors sensitive to large numbers of photons and which respond monotonically to incident power (such as photodiodes) are widespread; however, they exhibit significant electrical noise, which washes out any photon-number-resolving capability. 

Bridging the gap between single-photon-level measurements and bright light has significant implications for both fundamental studies of large-scale photonic quantum states, as well as practical implications for optical metrology. In this paper, we show that by combining intrinsic photon-number resolution and large-scale multiplexing, we not only bridge this gap, but do so whilst exceeding the precision of Poisson-noise-limited single-photon-level measurements. 

To achieve this, we use the spatially- and temporally-multiplexed detector shown in Fig.~\ref{fig:multiplexing}(a). This device comprises 128 temporal modes incident on eight intrinsically photon-number-resolving superconducting nanowire single-photon detectors (PNR-SNSPDs), resulting in 1024 multiplexing bins. Each bin is rigorously characterized by mapping its arrival time relative to a trigger signal to the probability a certain number of photons were absorbed, following the analysis in Refs.~\cite{sidorova2025jitter,schapeler2026practical}. For a given arrival time measured by the time tagger, we thus have a distribution of photon-number probabilities corresponding to that arrival time. In each shot of the experiment, we obtain 1024 individual samples of the incident light, each of which corresponds to a calibrated photon number distribution. These samples are convolved to yield an overall distribution of photon numbers, and thus an estimate of the photon number absorbed by the device. 

\section{Experimental Method}\label{sec:setup}
\begin{figure*}
    \centering
    \includegraphics[width=1\linewidth]{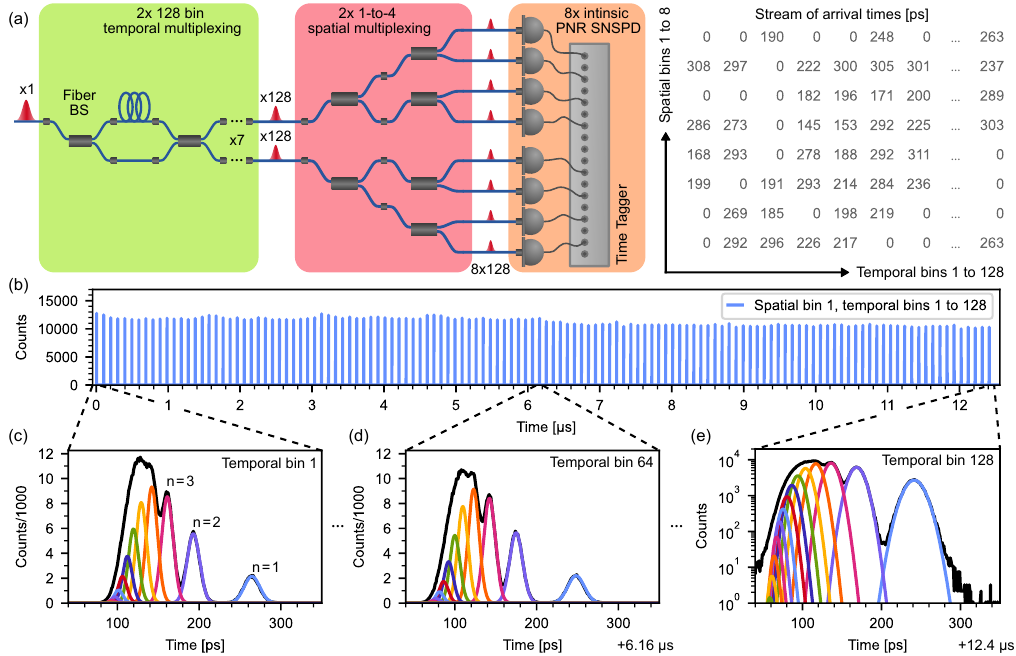}
    \caption{(a) Combination of temporal multiplexing (green box), spatial multiplexing (red box) and intrinsic PNR SNSPDs (orange box). One input pulse is (temporally) split into two pulse trains with 128 sub pulses each, afterwards both pulse trains are (spatially) split onto four SNSPDs, which can resolve the photon number in each of the $8\cdot128=1024$ resulting sub pulses, i.e, bins. The output of each of the eight spatial bins is a stream of 128 arrival times measured by the time tagger, which is depicted in the upper right corner. (b) Arrival-time histogram of the entire pulse train, i.e., all 128 temporal bins for the first spatial bin. The 128 peaks correspond to the individual temporal bins, which are spaced approximately $100~\mathrm{ns}$ apart for a total duration of $\approx12.5~\mathrm{\upmu s}$. Zooming into the histogram yields the arrival-time histograms of the individual temporal bins shown in (c), (d) and (e) for bins 1, 64 and 128, respectively (with (e) shown on a logarithmic scale). Here, for an incident mean photon number $\bar{n}=5$, the individual photon-number contributions for $n=1,2,3,...$ can be seen. The colored lines correspond to exponentially modified Gaussian (EMG) fits based on the model from Sidorova et al.~\cite{sidorova2025jitter}.}
    \label{fig:multiplexing}
\end{figure*}
 
Our experimental approach is shown in Fig.~\ref{fig:multiplexing}(a), which allows us to split an optical pulse into eight 128-bin pulse trains, resulting in a total of 1024 detection bins, each with intrinsic photon-number resolution.
Optical pulses enter a 128-bin-per-output temporal multiplexing network (green box). This fiber network consists of balanced beam splitters and delay lines (made from polarization maintaining components). The upper arms of the beam splitters are connected to delay lines with lengths according to $2^xL$, where $x\in[0,6]$ and $L=20~\mathrm{m}$, corresponding to a time delay of approximately $100~\mathrm{ns}$ (see Ref.~\cite{tiedau2020phdthesis} for more detail on the temporal multiplexing setup). For every impinging optical pulse, the network creates a pulse train consisting of 128 pulses for both outputs (shown for one spatial bin in Fig.~\ref{fig:multiplexing}(b)). This means that the intensity of the impinging pulse is split evenly across all 256 sub-pulses. 

Afterwards, the generated pulse trains enter a spatial multiplexing network (red box), which divides each input into four spatial bins using three balanced beam splitters. Every spatial bin is connected to a commercial photon-number-resolving SNSPD (Single Quantum). Finally, a time-to-digital converter (Time Tagger X, Swabian Instruments) records the raw time tags for later analysis (orange box).

\vspace{+1em}
We use PNR SNSPDs (based on arrival-time measurements), therefore, every bin can resolve the number of photons in the sub-pulses (shown in Fig.~\ref{fig:multiplexing}(c)-(e) for three temporal bins). In total, we measure 125 input states with mean photon numbers scaling quadratically from 0 to 15393 per pulse; we record $3.2\cdot10^6$ samples for each input state. The calibration of the coherent input states is performed following the procedure outlined in Refs.~\cite{schapeler2020quantum,schapeler2024electrical}, relating the incident mean photon number per pulse of the coherent states to the attenuation of the variable optical attenuators.
More details about the full experimental setup is presented in the supplemental document (see Sec.~1 and Fig.~1 therein).

\begin{figure*}[ht]
    \centering
    \includegraphics[width=1\linewidth]{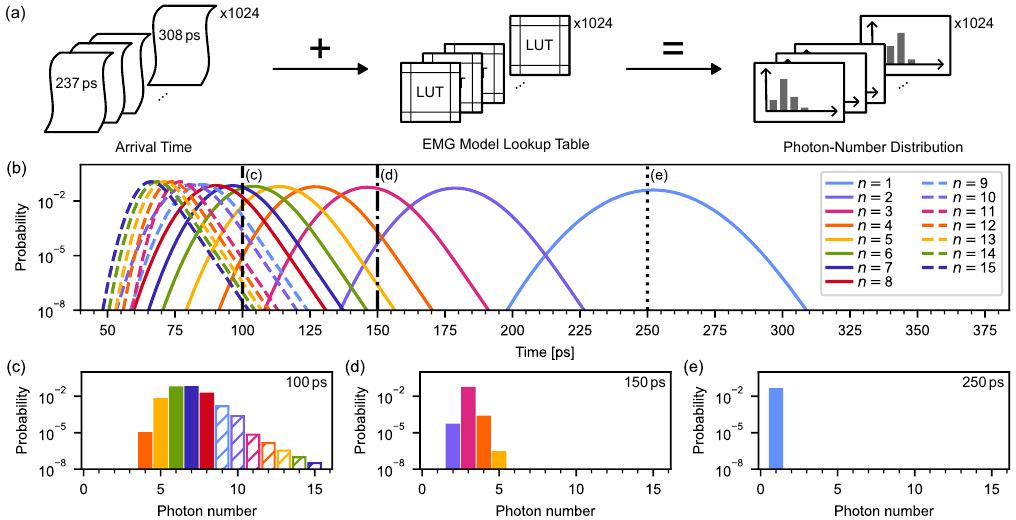}
    \caption{(a) For every input state that impinges on the detection system, we record up to 1024 arrival times (corresponding to the $8\cdot128=1024$ spatial and temporal bins). For every bin, we have a lookup table (LUT), which is created through an EMG model describing the $n=15$ underlying photon-number contributions (b). With this, every arrival time is then translated into a photon-number distribution (c)-(e). (b) The LUT is created by fitting EMG distributions to the arrival-time histograms (see Fig.~\ref{fig:multiplexing}(c)-(e)) for all 1024 bins and then assuming a flat prior distribution. As an example we show the LUT for spatial bin 1 and temporal bin 48. (c)-(e) The photon-number distributions are a result of a vertical cut at the arrival times highlighted in (b), i.e., $100~\mathrm{ps}$, $150~\mathrm{ps}$ and $250~\mathrm{ps}$, respectively, obtaining the probability values of the underlying distributions and renormalization.}
    \label{fig:LUT}
\end{figure*}

\section{Shot-to-shot photon-number distributions from constituent detectors}
In general, the mapping of a phase insensitive detector outcome to photon number is described by the positive operator valued measure (POVM) of the detector~\cite{lundeen2009tomography}. Before investigating the POVMs of the device in its entirety, it is helpful to consider the measurement outcomes arising from the constituent detection bins. As we have previously shown~\cite{sidorova2025jitter}, the rise-time of the electrical signal from SNSPDs depends on the number of hotspots generated within a characteristic time (typically $\lesssim 50~\mathrm{ps}$). Therefore, assuming these hotspots are caused by multiple photons, the detector becomes photon-number resolving. Moreover, since these hotspots are small compared to the total nanowire area, the probability of overlapping hotspots or two photons being absorbed within the same hotspot is negligible~\cite{sidorova2025jitter}. Thus the detector is considered photon-number resolving, up to the point where the photon-number dependent rise times cannot be distinguished (typically $\lesssim 5$ photons~\cite{cahall2017multi,zhu2020resolving,endo2021quantum,sauer2023resolving,schapeler2024electrical,kong2024large,los2024high,jaha2024kinetic,sidorova2025jitter,schapeler2026practical}).

\begin{figure*}[ht]
    \centering
    \includegraphics[width=1\linewidth]{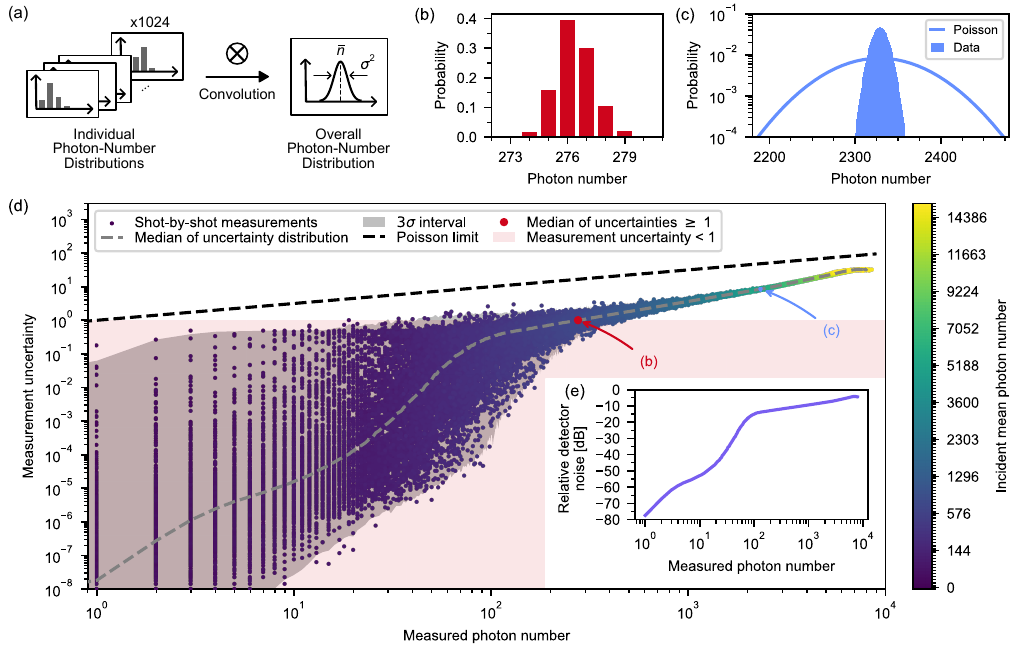}
    \caption{(a) From the 1024 individual photon-number distributions generated with the LUT (as depicted in Fig.~\ref{fig:LUT}(c)-(e)) we can generate an overall photon-number distribution for every single input state that impinges on the detection system. This is done via a convolution. (b), (c) Photon-number distribution of a single shot for two different incident mean photon numbers (484 and 4904, respectively). These shots are marked in (d) with red and blue dots, respectively. (d) Scatter plot of the measurement uncertainty vs. the measured photon number on a shot-by-shot basis. For every incident mean photon number (corresponding to different colors of the scatter plot dots, see color bar) we plot 1000 samples. It can be seen that all data points are below the Poisson noise limit (black dashed line), which corresponds to $\sigma=\sqrt{\bar{n}}$. For every measured photon number we find the distribution of measurement uncertainties and determine the median (gray dashed line) and $3\sigma$ interval (gray shaded area). The red shaded region indicates where the measurement uncertainty is below 1. The red dot indicates the measured photon number, where the median of the uncertainty distributions exceeds 1 for the first time, which happens for a measured photon number of 276. (e) Relative detector noise vs. measured photon number. The detector operates consistently below the Poisson noise limit, with a maximum relative detector noise of $-4.1~\mathrm{dB}$ at $>6800$ photons.}
    \label{fig:varVSmean}
\end{figure*}

\subsection{Arrival time to photon number mapping} \label{sec:luts}
In the experiment, we measure an arrival time, from which we wish to infer the photon number. Recently, Sidorova et al.~\cite{sidorova2025jitter} presented an analytical model, based on exponentially-modified Gaussian (EMG) distributions, which maps the number of absorbed photons to arrival times, taking into account various sources of jitter and finite timing resolution. Therefore, no unique mapping between arrival time and photon number is possible. Instead, for a given measured arrival time, there exists a distribution of contributions from different numbers of photons. This mapping can be seen as POVMs of each individual detection bin on a restricted Hilbert space; the mapping is only valid when a detection event occurs, i.e., the vacuum event is not covered by this restricted space. Assuming saturated internal detection efficiency, the model describes the behavior of the detector upon photon absorption, and is ignorant of any loss. Vacuum events are registered as no response from the detector during a measurement time window. This is incorporated into the mapping by the distribution $p(n)=\delta_{n,0}$, {i.e.,} the unit probability for the vacuum component and zero for all other photon numbers.

\newpage
Given a flat prior of input photon numbers (i.e., a uniform distribution, as not to assume any prior knowledge about the input photon statistics), one can use the Sidorova model to find this restricted POVM. In practice, this results in a lookup table (LUT), whereby each arrival time corresponds to a probability distribution of photon numbers (schematically shown in Fig.~\ref{fig:LUT}(a)). We truncate the model to include photon-number contributions (per bin) up to 15 (a lookup table for one detection bin is shown in Fig.~\ref{fig:LUT}(b)). This approach has the significant advantage that we incorporate and maintain the probabilistic contribution of all photon numbers throughout the analysis, which allows us to put error bars on each measurement shot-to-shot.

\begin{table*}[ht]
\caption{Comparison of recent works, showing the maximum photon number $n_\mathrm{max}$ with a $\pm1$ photon uncertainty ($\sigma\leq1$). $^\ast$Based on the LUTs for all bins, where the standard deviation exceeds one for the first time. $^\dagger$Values are calculated with Eq.~8 from Ref.~\cite{miatto2018explicit}, where the standard deviation of the click distribution is $\leq1$.$^\ddagger$ Inferred from the photon-number error rates in the extended data given in Ref.~\cite{eaton2023resolution}.}
\label{tab:comparison}
\begin{tabular}{l|l|l|c}
Source & Detector type & Methodology & $n_\mathrm{max}$ \\ \hline
This work & SNSPD & single-pixel & 8$^\ast$ \\
Stasi et al.~\cite{stasi2024enhanced} & SNSPD & 28-pixel interleaved & 10$^\dagger$  \\
Cheng et al.~\cite{cheng2023a} & SNSPD & 100-pixel waveguide-integrated & 16$^\dagger$ \\
This work & SNSPD & 1024-bin multiplexing (no PNR) & 47$^\dagger$ \\
Levine et al.~\cite{levine2014photon} & TES & Poisson-influenced $K$-means algorithm~\cite{levine2012algorithm} & 100 \\
Eaton et al.~\cite{eaton2023resolution} & TES & 3-bin spatial multiplexing & $\lesssim106^\ddagger$  \\
This work & SNSPD & 1024-bin multiplexing combined with intrinsic PNR & 276 \\
\end{tabular}
\end{table*}

\subsection{Photon-number distributions}
For each shot of the experiment, we generate 1024 distributions $p_i\left(n\right)$ corresponding to the probability that a photon number $n$ contributed to the outcome of each bin $i$ (shown for one bin for three different arrival times in Fig.~\ref{fig:LUT}(c)-(e)). To generate the aggregated distribution $P\left(n\right)$, we convolve all 1024 distributions $(p_1 \ast p_2 \ast \dots \ast p_N)\left(n\right) = \left( {\bigconv}_{i=1}^{B} p_i \right)\left(n\right)$, which is schematically shown in Fig.~\ref{fig:varVSmean}(a). This convolution is computationally efficient by multiplying each distribution in Fourier space and then taking the inverse Fourier transform of the result to obtain the final distribution. The result comprises the probability of each photon number contributing to the final outcome. The mean of the resulting distribution is the photon number, and the standard deviation its error, generated on a shot-by-shot basis. 

An example of a measurement outcome following this analysis is shown in Figs.~\ref{fig:varVSmean}(b),(c). As can be seen, this corresponds to measuring a mean of 276 and 2329 photons, respectively. The width of the distribution is significantly narrower than the Poisson noise limit (i.e., shot noise following Poissonian statistics) indicated by the solid line in Fig.~\ref{fig:varVSmean}(c). This corresponds to a significant improvement in precision compared to the Poisson noise limit of over an order of magnitude ($12.1~\mathrm{dB}$ and $7.5~\mathrm{dB}$, respectively).

While the above analysis maintains complete information about all photon-number contributions to the final distribution, this may not be necessary in some measurement tasks. In an alternative analysis, we can consider each of the 1024 bins as sampling the initial incident photon-number distribution. One can calculate the means $\bar{n}_i$ and variances $\sigma_i^2$ of each of these samples, and, since they are independent and therefore additive, sum them over the total number of bins $B=1024$ to yield the resulting mean $\bar{n}=\sum_i^B\bar{n}_i$ and variance $\sigma^2=\sum_i^B \sigma^2_i$ or standard deviation $\sigma=\sqrt{\sigma^2}$. This mean and variance are identical to the mean and variance that can be calculated from the previously discussed method, i.e., the complete photon-number distribution.

\begin{figure*}
    \centering
    \includegraphics[width=1\linewidth]{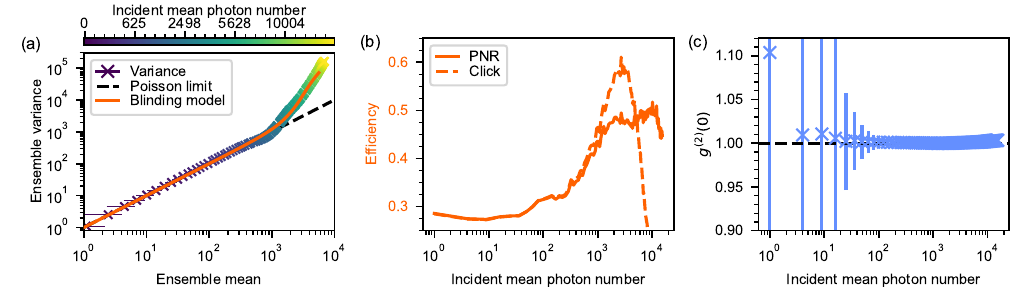}
    \caption{(a) Ensemble variance vs. ensemble mean from $10^5$ samples for every incident mean photon number, corresponding to the colored cross symbols. These lie on the Poisson noise limit (black dashed line) as we measure coherent light, which follows Poissonian statistics. Above a measured mean of approximately 1000 we see a deviation, which we can attribute to a probabilistic blinding of our detection bins. Our blinding model (orange line) accurately explains the deviation from the Poisson noise limit (see Sec~III~D in the supplemental document for more detail). (b) The efficiency of the device is the fraction of the measured mean photon number over the incident mean photon number (orange line). The dashed orange line corresponds to the efficiency when operating the detector in click-detection mode. (c) Second-order correlation function $g^{(2)}(0)$ as a function of incident mean photon number. The weighted average is $g^{(2)}(0)=1.000270(7)$.}
    \label{fig:ensemble_g2}
\end{figure*}

\subsection{Shot-by-shot measurement uncertainty vs. measured photon number}
We apply the alternative (second) method to measure a series of coherent states, ranging in incident mean photon number from 0 to 15393. In Fig.~\ref{fig:varVSmean}(d) we plot $10^3$ shot-by-shot data points (measurement uncertainty $\sigma$ vs. measured photon number $\bar{n}$) for each incident mean photon number (the colors of the dots correspond to different incident mean photon numbers, as shown in the color bar).

As stated above, these can be calculated directly from the aggregated distribution, or equivalently the sums of the variances or means of the 1024 constituent samples.
For a subset of our data set ($12.5\cdot10^6$ data points, i.e., $10^5$ shots per incident mean photon number), we aggregate the distribution of measurement uncertainties for all integer valued measured photon numbers (corresponding to the x axis in Fig.~\ref{fig:varVSmean}(d), from 1 to the largest measured photon number of approximately 9000). We indicate the two example shot-by-shot measurements from Figs.~\ref{fig:varVSmean}(b),(c) with dots and arrows in the same color as the respective plots. The median as well as the $3\sigma$ interval of these distributions are shown as the gray dashed line and the gray shaded area in Fig.~\ref{fig:varVSmean}(d).
To display the uncertainty, we use the median, i.e, the midpoint of the measurement uncertainty distribution for a certain measured photon number.

For Poisson noise limited measurements, the measured mean and the variance are equal, hence the measurement uncertainty scales as the square root of the mean. This limit is indicated by the black dashed line (corresponding to $\sigma=\sqrt{\bar{n}}$). The measurement uncertainties are consistently below the Poisson noise limit for all measured photon numbers. In Fig.~\ref{fig:varVSmean}(e) we show the relative detector noise $N_\mathrm{dB} = 10\cdot\mathrm{log}_{10}\left(\frac{\sigma}{\sqrt{\bar{n}}}\right)$ in $\mathrm{dB}$, which compares the measurement uncertainty to the Poisson noise limit. We find an average relative detector noise of $-6.7~\mathrm{dB}$ across the entire tested dynamic range of the detector, i.e., almost a factor of five below the Poisson noise limit. The maximum detector noise is $-4.1~\mathrm{dB}$ at $>6800$ photons and the minimum is $-77.3~\mathrm{dB}$ at $1$ photon. Note, at this point, we are limited by the dark-count probability of our detector, which is $<3.7\cdot10^{-5}$ based on a combined dark-count rate of $<80~\mathrm{Hz}$ of the eight SNSPD channels.

Measurement uncertainties below one, i.e., all data points in the red shaded region in Fig.~\ref{fig:varVSmean}(d), correspond to better than $\pm1$ photon precision in the measurements ($\sigma<1$), which we achieve up to a measured photon number of 276.
To give a sense of scale, we can calculate the number of click detectors one would need, to get a standard deviation of the click distribution of $\leq1$. Using Eq.~8 from Ref.~\cite{miatto2018explicit} this results in $3.7\cdot10^4$ click detectors, i.e., well over one order of magnitude more detectors than detection bins used in this work. Table~\ref{tab:comparison} compares the maximum photon number $n_\mathrm{max}$ with a $\pm1$ photon uncertainty; it shows that our detector significantly outperforms other published work.
This shows the power of combining multiplexing with intrinsic photon-number resolution, where intrinsically resolving the photon number is the crucial ingredient for the high precision, enabled by the large separation of the one-photon and ``more than one"-photon events in the arrival-time-based PNR measurements (compare Fig.~\ref{fig:multiplexing}(c)-(e)). 

\section{Photon-number resolved measurements of coherent states}
Here we discuss how ensemble data from the different coherent input states is used to further characterize the detector and our coherent light source. We investigate the variance and mean of ensemble data, estimate the efficiency of the detector, and calculate the second order correlation function using our photon-number resolved measurements.

\vspace{-1em}
\subsection{Ensemble data}\label{sec:ensembleData}
In Fig.~\ref{fig:ensemble_g2}(a) we show the ensemble variance over the ensemble mean of $10^5$ samples for every incident mean photon number of the experiment (colored cross symbols, where the color corresponds to different incident mean photon numbers according to the color bar). We expect these to follow the Poisson noise limit, since they are sampling the underlying Poissonian photon statistics of the coherent input states. This is the case up to a measured mean photon number of approximately 1000 photons. Beyond this point, the variance of these distributions increase significantly. This is due to blinding effects of the detector bins, which leads to a zero-inflated distribution, with a mean photon number dependent probability of a ``blinded'' detector bin. More details are discussed in Sec.~III~D in the supplemental document, where we developed a model (orange line in Fig.~\ref{fig:ensemble_g2}(a)) to explain this behavior.

\subsection{Estimating the efficiency}
We calculate the efficiency of the device as the fraction of the measured mean photon number and the incident mean photon number, which is shown in Fig.~\ref{fig:ensemble_g2}(b) as a function of the incident mean photon number (orange line). It is apparent that the efficiency has a strong nonlinear dependence on the incident power. We attribute this to the detector count rate dependent efficiency for AC-coupled amplification circuits~\cite{kerman2013readout,ferrari2019analysis,mark2024nonlinear}. Due to the temporal multiplexing architecture higher mean photon numbers lead to higher detector count rates, as more bins contain photons, that will lead to more detection events. More detail is given in Sec.~III~A in the supplemental document. Despite these effects, the detector can reach an efficiency of above 50\% for any incident mean photon number, when setting the appropriate operational bias current. For this data set the efficiency of optimized for an incident mean photon number of approximately 5000 photons per pulse.

As a point of comparison, we can treat the detector as a ``click'' detector, and estimate the mean photon number based on the probability of measuring vacuum (``no click'' events) in a $450~\mathrm{ps}$ coincidence window for each bin. Since we assume the incident light to be Poissonian, the probability of vacuum is $P\left(0,\bar{n}\right)=e^{-\bar{n}\eta}$ and the efficiency can be expressed as $\eta=-\mathrm{ln}(P_\mathrm{no\,click})\frac{1}{\bar{n}}$. This is also shown in Fig.~\ref{fig:ensemble_g2}(b) as the orange dashed line. Here again, the nonuniform efficiency is apparent. Furthermore, while the two estimates of the mean photon number agree up to an incident mean photon number of approximately 500, they deviate significantly beyond this point. This is due to noise-induced ``blinding'' of some detector bins, which increases the probability of measuring a zero-event, when the detector is under high illumination. This is discussed in more detail in Sec.~III~D in the supplemental document. For much higher incident mean photon numbers the efficiency 

The accuracy of the detector, namely how well it can infer the input photon number from the measurement outcomes, depends on how well the losses can be characterized. However, its underlying precision, {i.e.}, how precisely number of incident photons can be counted, depends on the variance of the measurements, which is independent of loss.

\subsection{Second order autocorrelation}
Given photon-number resolved measurements we can also calculate the second order correlation function from 
\begin{equation}
    g^{(2)}(0) = 1+\frac{\sigma^2-\bar{n}}{\bar{n}^2}\,,
\end{equation}
using the ensemble variance and ensemble mean (i.e., the cross symbols from Fig.~\ref{fig:ensemble_g2}(a)). We show the $g^{(2)}(0)$ values for all incident mean photon numbers in Fig.~\ref{fig:ensemble_g2}(c). The weighted average calculates to $\bar{g}^{(2)}(0)=1.000270(7)$. As expected, our light source produces coherent light across a broad range of mean photon numbers.

\begin{figure}
    \centering
    \includegraphics[width=1\linewidth]{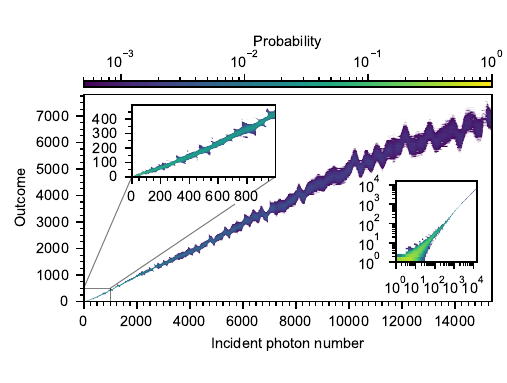}
    \caption{Reconstructed POVM matrix of the entire detector shown as a heatmap of the conditional probabilities. The x axis corresponds to the incident mean photon number and the y axis to the detector outcome, i.e., the measured photon number. The color axis shows the probability of observing the conditional probability of a certain measured photon number given an incident photon number. The top left inset shows a zoom-in of the first 500 outcomes. The bottom right inset shows the content of the main plot on a logarithmic scale for both axes.}
    \label{fig:POVMs}
\end{figure}

\section{Quantum Detector Tomography}\label{sec:povms}
The entire outcome space of the device is $1024^{450}$, {i.e.,} all combinations of each bin and each possible arrival time. This unfeasibly large number makes many approaches to detector tomography intractable, since the space cannot be sampled in reasonable time. 

We can reduce our outcome space significantly by attributing photon numbers to each outcome, given an input, truncated at some reasonable incident photon number. This is possible due to the model-based LUTs (see Sec.~\ref{sec:luts}), which map arrival times to photon-number distributions for all detection bins. Note that the assignment is limited by the fidelity of the model, which may lose accuracy under certain illumination regimes (see Sec.~III~C in the supplemental document).
Specifically, we populate the outcome matrix $\mat{P}_{D\times N}$, by creating a histogram (with a total of $N$ photon-number bins) of the measured photon numbers $n'\in[0,9010]$. This is done for an ensemble of $10^5$ sampless for every of the $D=125$ incident mean photon numbers (more detail about the outcome matrix can be found in Sec.~{II} and Fig.~2 in the supplemental document). The input state matrix $\mat{F}_{D\times M}$ corresponds to Poisson distributions of the different incident coherent states truncated at a maximum photon number of 15393.

The POVMs are commonly displayed in a matrix $\mat{\Pi}$ consisting of the diagonal elements of the underlying set of POVMs $\{\mat{\pi}\}$, which contains all relevant data for phase-insensitive detectors (which applies to our detector). Therefore, this matrix displays the conditional probabilities $p(n'|n)$ of measuring $n'$ photons given $n$ incident photons.
The size of the POVM matrix $\mat{\Pi}_{M\times N}$ is thus $M\cdot N=15394\cdot9011\approx1.38\cdot10^8$ elements. 
We reconstruct the POVM matrix using established methods, optimized for reconstructing large matrices~\cite{schapeler2025scalable}, which are shown in Fig.~\ref{fig:POVMs}. In general, it can be seen that the POVMs are narrow across the entire Hilbert space $M\in[0,15393]$ and follow a slope of approximately $0.5$, indicating a detection efficiency of approximately 50\%. Some numerical artifacts are visible, especially at large incident photon numbers, which are amplified due to the logarithmic color bar scaling. The top left inset shows a zoom-in on the first 500 outcomes, displaying a linear dependency from input to output photon number, particularly in this regime. The bottom right inset shows the POVMs on logarithmic axes. Here, it can be seen that the detector is not yet saturating, which means that the dynamic range of the detector is not being fully utilized by the incident optical power. The most probable outcome for the largest input state is approximately 7000 photons. At the repetition rate of the experiment of $80~\mathrm{kHz}$, this corresponds to an optical power of over $71~\mathrm{pW}$. As the dynamic range of the device was not fully utilized, we could measure up to $156~\mathrm{pW}$ of optical power per pulse, if all detection bins yield the maximal photon-number event of 15 (based on the current implementation of the EMG-model, i.e., LUTs).

\section{Conclusion}
As the size of photonic quantum systems increases, building and analyzing sufficiently large detection systems will become increasingly important. Building up larger systems from well-characterized constituent components is key to this task. We applied this approach to intrinsically photon-number-resolving detectors. Convolving the resulting outcome distributions of each detection bin allowed us to construct the response of the device as a whole. This, in turn, allowed us to perform quantum detector tomography on the complete device. We show sub-single-photon precision up to 276 photons, and measure up to $71~\mathrm{pW}$ of optical power per pulse.

Our results show that this approach can maintain \emph{precise} photon-number resolution when measuring thousands of photons. Nevertheless, our \emph{accuracy} is limited by detector imperfections, particularly effects caused by temporary blinding of certain detector bins. Nevertheless, we show that the principle of combining detectors, each with intrinsic PNR, vastly outperforms multiplexed click detectors alone. Furthermore, we do so at a scale that is an order of magnitude larger than previous multiplexed single-photon counting systems~\cite{cheng2023a,eaton2023resolution}.

\begin{acknowledgments}
The authors thank Kai-Hong Luo and Simone Atzeni for helpful discussions during the experiment, Niklas Lamberty for discussions during the modeling of the blinding effects, Martin J. Stevens and Samantha Davis for discussions about the measurement uncertainty and Robert Schade for modifications of pqdts~\cite{pqdts} (the quantum detector tomography solver).
This work was partially funded by the European Union (ERC, QuESADILLA, 101042399). Views and opinions expressed are however those of the author(s) only and do not necessarily reflect those of the European Union or the European Research Council Executive Agency. Neither the European Union nor the granting authority can be held responsible for them. 
This work has received funding from the German Federal Ministry of Research, Technology and Space within the PhoQuant project (grant number 13N16103). 
F.S. is part of the Max Planck School of Photonics supported by the Dieter Schwarz Foundation, the German Federal Ministry of Research, Technology and Space (BMFTR), and the Max Planck Society.
\end{acknowledgments}

\section*{Data Availability Statement}
The data that support the findings of this study are openly available in Zenodo at \url{https://doi.org/10.5281/zenodo.19678965}.

\bibliography{references}

@article{schapeler2024electrical,
author = {Schapeler, Timon and Lamberty, Niklas and Hummel, Thomas and Schlue, Fabian and Stefszky, Michael and Brecht, Benjamin and Silberhorn, Christine and Bartley, Tim J},
doi = {10.1103/PhysRevApplied.22.014024},
journal = {Physical Review Applied},
month = {jul},
number = {1},
pages = {14024},
publisher = {American Physical Society},
title = {{Electrical trace analysis of superconducting nanowire photon-number-resolving detectors}},
url = {https://link.aps.org/doi/10.1103/PhysRevApplied.22.014024},
volume = {22},
year = {2024}
}

@article{gerrits2012extending,
author = {Thomas Gerrits and Brice Calkins and Nathan Tomlin and Adriana E. Lita and Alan Migdall and Richard Mirin and Sae Woo Nam},
journal = {Opt. Express},
number = {21},
pages = {23798--23810},
publisher = {Optica Publishing Group},
title = {Extending single-photon optimized superconducting transition edge sensors beyond the single-photon counting regime},
volume = {20},
month = {Oct},
year = {2012},
url = {https://opg.optica.org/oe/abstract.cfm?URI=oe-20-21-23798},
doi = {10.1364/OE.20.023798},
}

@article{schapeler2025scalable,
author = {Schapeler, Timon and Schade, Robert and Lass, Michael and Plessl, Christian and Bartley, Tim J},
doi = {10.1088/2058-9565/ad8511},
issn = {2058-9565},
journal = {Quantum Science and Technology},
month = {jan},
number = {1},
pages = {015018},
publisher = {IOP Publishing},
title = {{Scalable quantum detector tomography by high-performance computing}},
url = {https://dx.doi.org/10.1088/2058-9565/ad8511 https://iopscience.iop.org/article/10.1088/2058-9565/ad8511},
volume = {10},
year = {2025}
}

@article{lundeen2009tomography,
author = {Lundeen, J S and Feito, A and Coldenstrodt-Ronge, H and Pregnell, K L and Silberhorn, Ch. and Ralph, T C and Eisert, J and Plenio, M B and Walmsley, I A},
doi = {10.1038/nphys1133},
issn = {1745-2481},
journal = {Nature Physics},
number = {1},
pages = {27--30},
title = {{Tomography of quantum detectors}},
url = {https://doi.org/10.1038/nphys1133},
volume = {5},
year = {2009}
}

@article{sidorova2025jitter,
author = {Sidorova, Mariia and Schapeler, Timon and Semenov, Alexej D. and Schlue, Fabian and Stefszky, Michael and Brecht, Benjamin and Silberhorn, Christine and Bartley, Tim J.},
doi = {10.1063/5.0273752},
issn = {2378-0967},
journal = {APL Photonics},
month = {aug},
number = {8},
title = {{Jitter in photon-number-resolved detection by superconducting nanowires}},
url = {https://pubs.aip.org/app/article/10/8/086113/3360851/Jitter-in-photon-number-resolved-detection-by},
volume = {10},
year = {2025}
}

@article{eaton2023resolution,
author = {Eaton, Miller and Hossameldin, Amr and Birrittella, Richard J and Alsing, Paul M and Gerry, Christopher C and Dong, Hai and Cuevas, Chris and Pfister, Olivier},
doi = {10.1038/s41566-022-01105-9},
issn = {1749-4885},
journal = {Nature Photonics},
month = {jan},
number = {1},
pages = {106--111},
title = {{Resolution of 100 photons and quantum generation of unbiased random numbers}},
url = {https://doi.org/10.1038/s41566-022-01105-9 https://www.nature.com/articles/s41566-022-01105-9},
volume = {17},
year = {2023}
}

@article{schapeler2020quantum,
author = {Schapeler, Timon and {Philipp H{\"{o}}pker}, Jan and Bartley, Tim J},
doi = {10.1364/OE.404285},
journal = {Optics Express},
number = {22},
pages = {33035--33043},
publisher = {OSA},
title = {{Quantum detector tomography of a 2$\times$2 multi-pixel array of superconducting nanowire single photon detectors}},
url = {http://www.opticsexpress.org/abstract.cfm?URI=oe-28-22-33035},
volume = {28},
year = {2020}
}

@article{stasi2024enhanced,
author = {Stasi, Lorenzo and Taher, Towsif and Resta, Giovanni V and Zbinden, Hugo and Thew, Rob and Bussi{\`{e}}res, F{\'{e}}lix},
doi = {10.1021/acsphotonics.4c01680},
issn = {2330-4022},
journal = {ACS Photonics},
month = {jan},
number = {1},
pages = {320--329},
publisher = {American Chemical Society},
title = {{Enhanced Detection Rate and High Photon-Number Efficiencies with a Scalable Parallel SNSPD}},
url = {https://doi.org/10.1021/acsphotonics.4c01680 https://pubs.acs.org/doi/10.1021/acsphotonics.4c01680},
volume = {12},
year = {2025}
}

@article{los2024high,
author = {Los, J W Niels and Sidorova, Mariia and Lopez-Rodriguez, Bruno and Qualm, Patrick and Chang, Jin and Steinhauer, Stephan and Zwiller, Val and Zadeh, Iman Esmaeil},
doi = {10.1063/5.0204340},
issn = {2378-0967},
journal = {APL Photonics},
number = {6},
title = {{High-performance photon number resolving detectors for 850–950 nm wavelength range}},
url = {https://pubs.aip.org/app/article/9/6/066101/3296058/High-performance-photon-number-resolving-detectors},
volume = {9},
year = {2024}
}

@article{cahall2017multi,
author = {Cahall, Clinton and Nicolich, Kathryn L and Islam, Nurul T and Lafyatis, Gregory P and Miller, Aaron J and Gauthier, Daniel J and Kim, Jungsang},
doi = {10.1364/OPTICA.4.001534},
journal = {Optica},
number = {12},
pages = {1534--1535},
publisher = {OSA},
title = {{Multi-photon detection using a conventional superconducting nanowire single-photon detector}},
url = {http://www.osapublishing.org/optica/abstract.cfm?URI=optica-4-12-1534},
volume = {4},
year = {2017}
}

@article{sauer2023resolving,
archivePrefix = {arXiv},
arxivId = {quant-ph/2310.12472},
author = {Sauer, Gregor and Kolarczik, Mirco and Gomez, Rodrigo and Conrad, Johanna and Steinlechner, Fabian},
eprint = {2310.12472},
primaryClass = {quant-ph},
title = {{Resolving Photon Numbers Using Ultra-High-Resolution Timing of a Single Low-Jitter Superconducting Nanowire Detector}},
year = {2023},
journal = {arXiv preprint arXiv:2310.12472}
}

@article{endo2021quantum,
author = {Endo, Mamoru and Sonoyama, Tatsuki and Matsuyama, Mikihisa and Okamoto, Fumiya and Miki, Shigehito and Yabuno, Masahiro and China, Fumihiro and Terai, Hirotaka and Furusawa, Akira},
doi = {10.1364/OE.423142},
issn = {1094-4087},
journal = {Optics Express},
number = {8},
pages = {11728},
title = {{Quantum detector tomography of a superconducting nanostrip photon-number-resolving detector}},
url = {https://opg.optica.org/abstract.cfm?URI=oe-29-8-11728},
volume = {29},
year = {2021}
}

@article{kong2024large,
author = {Kong, Ling-Dong and Zhang, Tian-Zhu and Liu, Xiao-Yu and Li, Hao and Wang, Zhen and Xie, Xiao-Ming and You, Li-Xing},
doi = {10.1117/1.AP.6.1.016004},
journal = {Advanced Photonics},
number = {1},
pages = {16004},
title = {{Large-inductance superconducting microstrip photon detector enabling 10 photon-number resolution}},
url = {https://doi.org/10.1117/1.AP.6.1.016004},
volume = {6},
year = {2024}
}

@article{zhu2020resolving,
author = {Zhu, Di and Colangelo, Marco and Chen, Changchen and Korzh, Boris A and Wong, Franco N C and Shaw, Matthew D and Berggren, Karl K},
doi = {10.1021/acs.nanolett.0c00985},
issn = {1530-6984},
journal = {Nano Letters},
number = {5},
pages = {3858--3863},
publisher = {American Chemical Society},
title = {{Resolving Photon Numbers Using a Superconducting Nanowire with Impedance-Matching Taper}},
url = {https://doi.org/10.1021/acs.nanolett.0c00985},
volume = {20},
year = {2020}
}

@article{jaha2024kinetic,
author = {Jaha, Roland and Graham-Scott, Connor A and Abazi, Adrian S and Pernice, Wolfram and Schuck, Carsten and Ferrari, Simone},
doi = {10.1364/OE.564311},
issn = {1094-4087},
journal = {Optics Express},
month = {oct},
number = {20},
pages = {41869},
publisher = {Optica Publishing Group},
title = {{Kinetic inductance and jitter dependence of intrinsic photon-number resolution in superconducting nanowire single-photon detectors}},
url = {https://opg.optica.org/oe/abstract.cfm?URI=oe-33-20-41869 https://opg.optica.org/abstract.cfm?URI=oe-33-20-41869},
volume = {33},
year = {2025}
}

@article{lita2008counting,
author = {Lita, Adriana E and Miller, Aaron J and Nam, Sae Woo},
doi = {10.1364/OE.16.003032},
issn = {1094-4087},
journal = {Optics Express},
number = {5},
pages = {3032},
publisher = {OSA},
title = {{Counting near-infrared single-photons with 95\% efficiency}},
url = {http://www.osapublishing.org/oe/abstract.cfm?URI=oe-16-5-3032 https://opg.optica.org/oe/abstract.cfm?uri=oe-16-5-3032},
volume = {16},
year = {2008}
}

@article{cheng2023a,
author = {Cheng, Risheng and Zhou, Yiyu and Wang, Sihao and Shen, Mohan and Taher, Towsif and Tang, Hong X},
doi = {10.1038/s41566-022-01119-3},
issn = {1749-4885},
journal = {Nature Photonics},
month = {jan},
number = {1},
pages = {112--119},
title = {{A 100-pixel photon-number-resolving detector unveiling photon statistics}},
url = {https://doi.org/10.1038/s41566-022-01119-3 https://www.nature.com/articles/s41566-022-01119-3},
volume = {17},
year = {2023}
}

@article{harder2016single-mode,
  title = {Single-Mode Parametric-Down-Conversion States with 50 Photons as a Source for Mesoscopic Quantum Optics},
  author = {Harder, Georg and Bartley, Tim J. and Lita, Adriana E. and Nam, Sae Woo and Gerrits, Thomas and Silberhorn, Christine},
  journal = {Phys. Rev. Lett.},
  volume = {116},
  issue = {14},
  pages = {143601},
  numpages = {5},
  year = {2016},
  month = {Apr},
  publisher = {American Physical Society},
  doi = {10.1103/PhysRevLett.116.143601},
  url = {https://link.aps.org/doi/10.1103/PhysRevLett.116.143601}
}

@article{sperling2012true,
author = {Sperling, J and Vogel, W and Agarwal, G S},
doi = {10.1103/PhysRevA.85.023820},
journal = {Physical Review A},
month = {feb},
number = {2},
pages = {23820},
publisher = {American Physical Society},
title = {{True photocounting statistics of multiple on-off detectors}},
url = {https://link.aps.org/doi/10.1103/PhysRevA.85.023820},
volume = {85},
year = {2012}
}

@article{paul1996photon,
author = {Paul, H and T{\"{o}}rm{\"{a}}, P and Kiss, T and Jex, I},
doi = {10.1103/PhysRevLett.76.2464},
issn = {0031-9007},
journal = {Physical Review Letters},
month = {apr},
number = {14},
pages = {2464--2467},
publisher = {American Physical Society},
title = {{Photon Chopping: New Way to Measure the Quantum State of Light}},
url = {https://link.aps.org/doi/10.1103/PhysRevLett.76.2464},
volume = {76},
year = {1996}
}

@article{fitch2003photon,
author = {Fitch, M J and Jacobs, B C and Pittman, T B and Franson, J D},
doi = {10.1103/PhysRevA.68.043814},
journal = {Physical Review A},
month = {oct},
number = {4},
pages = {43814},
publisher = {American Physical Society},
title = {{Photon-number resolution using time-multiplexed single-photon detectors}},
url = {https://link.aps.org/doi/10.1103/PhysRevA.68.043814},
volume = {68},
year = {2003}
}

@article{achilles2003fiber,
author = {Achilles, Daryl and Silberhorn, Christine and {\'{S}}liwa, Cezary and Banaszek, Konrad and Walmsley, Ian A},
doi = {10.1364/OL.28.002387},
journal = {Optics Letters},
number = {23},
pages = {2387--2389},
publisher = {OSA},
title = {{Fiber-assisted detection with photon number resolution}},
url = {http://www.osapublishing.org/ol/abstract.cfm?URI=ol-28-23-2387},
volume = {28},
year = {2003}
}

@article{miatto2018explicit,
author = {Miatto, Filippo M and Safari, Akbar and Boyd, Robert W},
doi = {10.1364/AO.57.006750},
journal = {Applied Optics},
number = {23},
pages = {6750--6754},
publisher = {OSA},
title = {{Explicit formulas for photon number discrimination with on/off detectors}},
url = {http://www.osapublishing.org/ao/abstract.cfm?URI=ao-57-23-6750},
volume = {57},
year = {2018}
}

@phdthesis{tiedau2020phdthesis,
author = {Tiedau, Johannes},
doi = {10.17619/UNIPB/1-1031},
pages = {138},
school = {Paderborn University},
title = {{Quantum optics in the photon number basis}},
type = {{PhD Thesis}},
year = {2020}
}

@article{schapeler2026practical,
author = {Schapeler, Timon and Mischke, Isabell and Schlue, Fabian and Stefszky, Michael and Brecht, Benjamin and Silberhorn, Christine and Bartley, Tim J.},
doi = {10.1063/5.0304127},
issn = {2835-0103},
journal = {APL Quantum},
month = {mar},
number = {1},
title = {{Practical considerations for assignment of photon numbers with SNSPDs}},
url = {https://pubs.aip.org/apq/article/3/1/016102/3375946/Practical-considerations-for-assignment-of-photon},
volume = {3},
year = {2026}
}

@article{kerman2013readout,
author = {Kerman, Andrew J and Rosenberg, Danna and Molnar, Richard J and Dauler, Eric A},
doi = {10.1063/1.4799397},
issn = {0021-8979},
journal = {Journal of Applied Physics},
month = {apr},
number = {14},
pages = {144511},
title = {{Readout of superconducting nanowire single-photon detectors at high count rates}},
url = {https://doi.org/10.1063/1.4799397},
volume = {113},
year = {2013}
}

@article{mark2024nonlinear,
author = {Mark, Patrick and Gstir, Sebastian and M{\"{u}}nzberg, Julian and Weihs, Gregor and Keil, Robert},
doi = {10.1063/5.0231297},
issn = {2835-0103},
journal = {APL Quantum},
month = {nov},
number = {4},
pages = {46109},
title = {{Nonlinear response of telecom-wavelength superconducting single-photon detectors}},
url = {https://doi.org/10.1063/5.0231297},
volume = {1},
year = {2024}
}

@article{ferrari2019analysis,
author = {Ferrari, Simone and Kovalyuk, Vadim and Vetter, Andreas and Lee, Changhyoup and Rockstuhl, Carsten and Semenov, Alexander and Gol'tsman, Gregory and Pernice, Wolfram},
doi = {10.1063/1.5113652},
issn = {0003-6951},
journal = {Applied Physics Letters},
month = {sep},
number = {10},
pages = {101104},
title = {{Analysis of the detection response of waveguide-integrated superconducting nanowire single-photon detectors at high count rate}},
url = {https://doi.org/10.1063/1.5113652},
volume = {115},
year = {2019}
}

@book{migdall2013single,
author = {Migdall, A and Polyakov, S V and Fan, J and Bienfang, J C},
isbn = {9780123876966},
publisher = {Academic Press},
title = {{Single-Photon Generation and Detection: Physics and Applications}},
year = {2013}
}

@software{pqdts,
author = {Schade, Robert and Lass, Michael and Schapeler, Timon and Plessl, Christian and Bartley, Tim J.},
title = {Parallel Quantum Detector Tomography Solver (pqdts)},
month = mar,
year = 2024,
publisher = {Zenodo},
version = {v2024.2},
doi = {10.5281/zenodo.10908474},
url = {https://github.com/pc2/pqdts}
}

@article{levine2014photon,
author = {Levine, Zachary H and Glebov, Boris L and Migdall, Alan L and Gerrits, Thomas and Calkins, Brice and Lita, Adriana E and Nam, Sae Woo},
doi = {10.1364/JOSAB.31.000B20},
journal = {Journal of the Optical Society of America B},
number = {10},
pages = {B20--B24},
publisher = {Optica Publishing Group},
title = {{Photon-number uncertainty in a superconducting transition edge sensor beyond resolved-photon-number determination}},
url = {https://opg.optica.org/josab/abstract.cfm?URI=josab-31-10-B20},
volume = {31},
year = {2014}
}

@article{levine2012algorithm,
author = {Levine, Zachary H and Gerrits, Thomas and Migdall, Alan L and Samarov, Daniel V and Calkins, Brice and Lita, Adriana E and Nam, Sae Woo},
doi = {10.1364/JOSAB.29.002066},
issn = {0740-3224},
journal = {Journal of the Optical Society of America B},
month = {aug},
number = {8},
pages = {2066},
publisher = {Optica Publishing Group},
title = {{Algorithm for finding clusters with a known distribution and its application to photon-number resolution using a superconducting transition-edge sensor}},
url = {https://opg.optica.org/josab/abstract.cfm?URI=josab-29-8-2066 https://opg.optica.org/abstract.cfm?URI=josab-29-8-2066},
volume = {29},
year = {2012}
}

\onecolumngrid
\clearpage
\foreach \i in {1,2,3,4,5,6}{
    \begin{figure*}[p]
        \thispagestyle{empty}
        \vspace*{-2cm} 
        \makebox[\textwidth][c]{ 
            \includegraphics[page=\i, width=1.00\paperwidth]{arXivSupplemental.pdf}
        }
    \end{figure*}
    \clearpage
}
\thispagestyle{empty}
\end{document}